\documentclass[[prb,preprint,showpacs,floatfix,amsmath,amssymb,superscriptaddress]{revtex4}

\usepackage[utf8]{inputenc}
\usepackage[T1]{fontenc}
\usepackage{amsmath}
\usepackage{amssymb,amsfonts,textcomp}
\usepackage{graphicx}
\usepackage[dvipsnames]{xcolor}
\usepackage{hyperref}

\newcommand{\Figref}[1]{Fig.~\ref{#1}}

\begin{document}

\title{Length-independent quantum transport through \\topological band states of graphene nanoribbons}

\author{
Song Jiang$^{1}$, Fabrice Scheurer$^{1}$, Qiang Sun$^{2,\dagger}$, Pascal Ruffieux$^{2}$, Xuelin Yao$^{3,\dagger\dagger}$, Akimitsu Narita$^{3}$, Klaus Müllen$^{3,4}$, Roman Fasel$^{2,5}$, Thomas Frederiksen$^{\ast,6,7}$, Guillaume Schull$^{\ast,1}$\\
\bigskip
\small{$^1$ Universit\'e de Strasbourg, CNRS, IPCMS, UMR 7504, F-67000 Strasbourg, France}\\
\small{$^2$ Empa - Swiss Federal Laboratories for Materials Science and Technology, nanotech@surfaces Laboratory, Überlandstrasse 129, 8600 Dübendorf, Switzerland}\\
\small{$^3$ Max Planck Institute for Polymer Research, Mainz, Germany}\\
\small{$^4$ Department of Chemistry Johannes Gutenberg University Mainz Duesbergweg 10-14 55128 Mainz Germany}\\
\small{$^5$ Department of Chemistry, Biochemistry and Pharmaceutical Sciences, University of Bern, Switzerland}\\
\small{$^6$ Donostia International Physics Center (DIPC) -- UPV/EHU, 20018 San Sebasti\'an, Spain}\\
\small{$^7$ IKERBASQUE, Basque Foundation for Science, 48013, Bilbao, Spain}\\
\small{$^\dagger$ Current address: Materials Genome Institute, Shanghai University, 200444 Shanghai, China}\\
\small{$^{\dagger\dagger}$ Current address: Department of Materials, University of Oxford, Oxford, UK}\\
\small{$^{\ast}$ email: thomas\_frederiksen@ehu.eus}\\
\small{$^{\ast}$ email: guillaume.schull@ipcms.unistra.fr}
}

\date{\today}


\maketitle

\textbf{In the realm of molecular electronics the active elements of an electronic device are composed of individual molecules \cite{McCreery2013Critical, metzger2015unimolecular, xiang2016molecular}. Among the requested components in this field, a substantial effort has been devoted to the design of linear organic systems, often referred to as molecular wires, capable of transporting large electrical currents with negligible losses over long distances. Despite many efforts from communities working either at room temperature in liquid environments \cite{xu2003measurement, tuccitto2009highly, Kolivoska2013single, carini2017high, yao2020long, Liang2022Highly, Lee2021Achieving} or at low temperature in ultra-high vacuum \cite{Lafferentz2009,  Reecht2014, Reecht2015,Nacci2015, skidin2018}, such a perfect molecular conductor has remained elusive. Graphene nanoribbons (GNRs), quasi-one-dimensional narrow strips of graphene, have emerged as promising candidates for high-performance molecular wires due to their extreme robustness inherited from graphene and their tunable energy band gaps resulting from lateral quantum confinement and edge effects \cite{Cai2010, wang2021graphene}. In this framework, special attention has been paid to the topological properties of specifically designed GNRs in both theoretical and experimental studies \cite{cao2017topological, groning2018engineering, rizzo2018topological, rizzo2020inducing}. These exotic properties rely on specific edge modifications of the ribbons that may open a new way to optimize the transport efficiency of GNRs \cite{koch2012voltage, jacobse2017electronic, jacobse2018mapping, lawrence2020probing, Mangnus2022Charge, Friedrich2022}. Here we present a systematic study of the charge transport properties of such topologically engineered GNRs, more specifically staggered edge-extended ribbons based on a 7-AGNR backbone denoted 7-AGNR-\textit{S}(1,3) \cite{groning2018engineering}, using a low-temperature scanning tunneling microscopy (LT-STM) lifting procedure. Our experiments reveal (quasi-) lossless transport properties at low voltages with high conductance over long distances. These exceptional properties arise from charge transport channels between the modified ribbon edges and are due to low-energy electronic topological bands originating from coupled localized zero energy states. We thus realized a longstanding dream of molecular electronics: length-independent charge transport through a molecular conductor.}

\clearpage
The transport properties of molecular wires were first investigated in solution, with molecular wires of increasing length bridging metallic electrodes \cite{xu2003measurement, su2016chemical, Xin2019}. In such an approach, the conductance--length relationship generally follows the simple exponential form $G(z) = G_c  \exp(-\beta z)$, where $z$ is the wire length, $G_c$ is the contact conductance and $\beta$, the inverse decay length which characterizes the intrinsic ability of the wire to transport the current \cite{Samanta1996, Khoo2015}. The lower the value of $\beta$, the more efficient is the transport of electrons through the wire. This approach was extensively used to probe soluble molecular wires, such as alkane or short conjugated structures, and has led, in a few cases, to $\beta$ values lower than 0.02 {\AA$^{-1}$} \cite{tuccitto2009highly, Kolivoska2013single, carini2017high, yao2020long, Liang2022Highly, Lee2021Achieving}. The recent development of on-surface synthesis (OSS) has considerably increased the types and lengths of molecular wires that can be probed \cite{clair2019controlling}.
Combined with LT-STM, it has led to a new type of experiment where the conductance of a given wire can be explored as a function of its length by progressively lifting the wire -- with one end attached to the STM tip -- from a metallic surface. 
The extremely well-controlled environment of this kind of experiment has allowed probing transport through defect-free wires several tens of nanometers long, and has revealed $\beta\sim0.1$-0.4 {\AA$^{-1}$} for conjugated polymers such as polyfluorene \cite{Lafferentz2009}, polythiophene \cite{Reecht2014,Reecht2015} and more complex donor-acceptor systems \cite{Nacci2015,skidin2018}. In this framework, GNRs have emerged as promising candidates for high-performance molecular wires \cite{wang2021graphene}.
Atomically precise fabrication of various types of GNRs has been achieved by OSS \cite{Cai2010, Narita2015New} and revealed fascinating electronic \cite{Hao2020Tuning, Houtsma2021Atomically}, magnetic \cite{Song2021On},  optical \cite{chong2018bright, Miao2021The}, and mechanical \cite{kawai2016superlubricity, koch2018structural} properties.
Transport measurements with a 7-atom-wide armchair GNR (7-AGNR), the most conventional GNR with armchair edge structure, were reported early on, revealing $\beta\sim 0.5$ {\AA$^{-1}$} for low-voltage conditions and $\beta \sim 0.1$ {\AA$^{-1}$} at higher voltages corresponding to resonant conditions \cite{koch2012voltage, jacobse2018mapping}. As such, the transport properties of 7-AGNRs are not significantly better than those of conjugated polymers, an effect that can be associated to their large electronic band gap ($\Delta >2.7$ eV) \cite{koch2012voltage}. Indeed, similar measurements on the quasi-metallic 5-AGNR revealed $\beta \sim 0.1$ {\AA$^{-1}$} even for low-voltage conditions \cite{lawrence2020probing}, and the transport properties of GNRs can be further tuned by mixing short 5- and 7-AGNR segments together \cite{jacobse2017electronic, Mangnus2022Charge}. 
Narrow GNRs doped with substitutional boron atoms have also been observed to exhibit a metallic response over up to 2.2 nm \cite{Friedrich2022}.
In the present study, we report on (nearly) lossless quantum transport through individual GNRs over long distances ($z > 10$ nm), corresponding to $\beta$ values two orders of magnitude lower than in previous studies ($\beta < 0.001$ {\AA$^{-1}$}), under low voltage condition (50 mV) while preserving  and high conductance ($\sim 0.1$ G$_0$, where G$_0 = 2e^2/h = 77.5$ $\mu$S is the quantum of conductance).

\begin{figure*}
\includegraphics[width=1\linewidth]{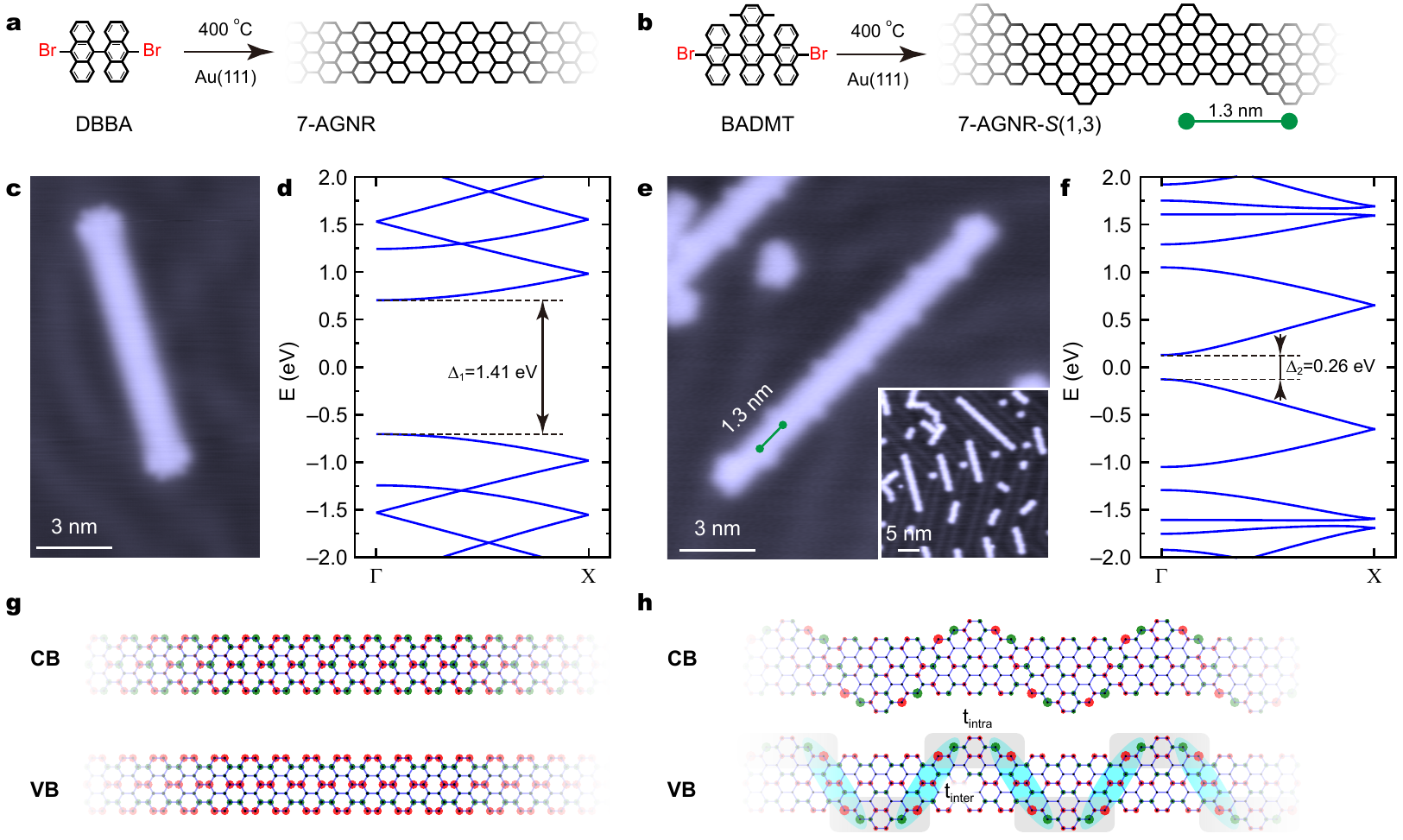}
\caption{\label{fig1}\textbf{Bottom-up synthesis of 7-AGNR and 7-AGNR-\textit{S}(1,3) on Au(111)}.
\textbf{a}, On-surface synthesis route of a 7-AGNR with precursor DBBA.
\textbf{b}, Synthesis route of a 7-AGNR-\textit{S}(1,3) with precursor BADMT.
\textbf{c}, STM image of a 7-AGNR on Au(111) ($V = 50$ mV, $I = 10$ pA).
\textbf{d}, Electronic band structure of a 7-AGNR from TB calculations in a 6-fold repeated cell (for direct comparison with panel \textbf{f}).
\textbf{e}, STM image of a 7-AGNR-\textit{S}(1,3) on Au(111) ($V = 50$ mV, $I = 10$ pA). The inset shows a large-scale STM image of a representative collection of 7-AGNR-\textit{S}(1,3) ribbons on Au(111).
\textbf{f}, Electronic band structure of 7-AGNR-\textit{S}(1,3) from TB calculations.
\textbf{g-h}, Wave functions of the conduction band (CB) and valence band (VB) at $\Gamma$ for a 7-AGNR (\textbf{g}) and a 7-AGNR-\textit{S}(1,3) (\textbf{h}), respectively.
Red and green denote opposite phases of the wave function. The low-energy bands in \textbf{h} can be described by an effective SSH model in terms of hopping integrals $t_\mathrm{intra}$ and $t_\mathrm{inter}$ as schematically indicated for the VB state.
}
\end{figure*}
To that end, we investigated the edge-extended 7-AGNR-\textit{S}(1,3)s \cite{groning2018engineering} and compared it to the conventional 7-AGNRs \cite{Cai2010}. Electronic transport through a lifted 7-AGNR has been explored thoroughly \cite{koch2012voltage, jacobse2018mapping} and is used here as a reference. 
These GNRs were assembled by OSS on Au(111) using 10,10'-dibromo-9,9'-bianthryl (DBBA, Figure \ref{fig1}a) \cite{Cai2010} and  6,11-bis(10-bromoanthracen-9-yl)-1,4-dimethyltetracene (BADMT, Figure \ref{fig1}b) \cite{groning2018engineering} as precursors for 7-AGNRs and 7-AGNR-\textit{S}(1,3)s, respectively. Figures \ref{fig1}c and \ref{fig1}e show typical STM images of these GNRs. The 7-AGNRs appear as featureless stripes with finger-like termini at small bias voltages (Figure \ref{fig1}c), consistent with previous reports \cite{koch2012voltage, talirz2013termini}. In contrast, the 7-AGNR-\textit{S}(1,3)s reveal protrusions alternating on both sides of the backbone with a half-periodicity of 1.3 nm (Figure \ref{fig1}e). The strict alternation is due to steric effects during the polymerization procedure \cite{groning2018engineering}. As shown in the large-scale image (inset of Figure \ref{fig1}e), the maximum length of the 7-AGNR-\textit{S}(1,3)s is approximately 20 nm, a limit that may be related to the typical dimensions of the Au(111) herringbone reconstruction \cite{Cai2010}.

The electronic band structures from simple tight-binding (TB) calculations of 7-AGNRs and 7-AGNR-\textit{S}(1,3)s, shown in Figure \ref{fig1}d and \ref{fig1}f, reveal band gaps of $\Delta_1 =1.41$ eV and $\Delta_2 =0.26$ eV, respectively, separating the valence band (VB) and conduction band (CB) states (Figure \ref{fig1}g for 7-AGNRs and Figure \ref{fig1}h for 7-AGNR-\textit{S}(1,3)s). This substantial reduction of the band gap upon edge decoration plays an important role for electron transport through the ribbons. Additionally, in contrast to the 7-AGNRs, the VB and CB states of 7-AGNR-\textit{S}(1,3)s are highly localized at the edge sites.  As reported in Ref.~\onlinecite{groning2018engineering} these low-energy bands can be rationalized from an SSH model with two hopping parameters along and across the backbone of nearly equal amplitude (t$_\mathrm{intra}$ = 0.45 eV, t$_\mathrm{inter}$ = 0.59 eV), resulting in a dispersive band with very small intrinsic band gap.

\begin{figure}
\includegraphics[width=0.5\linewidth]{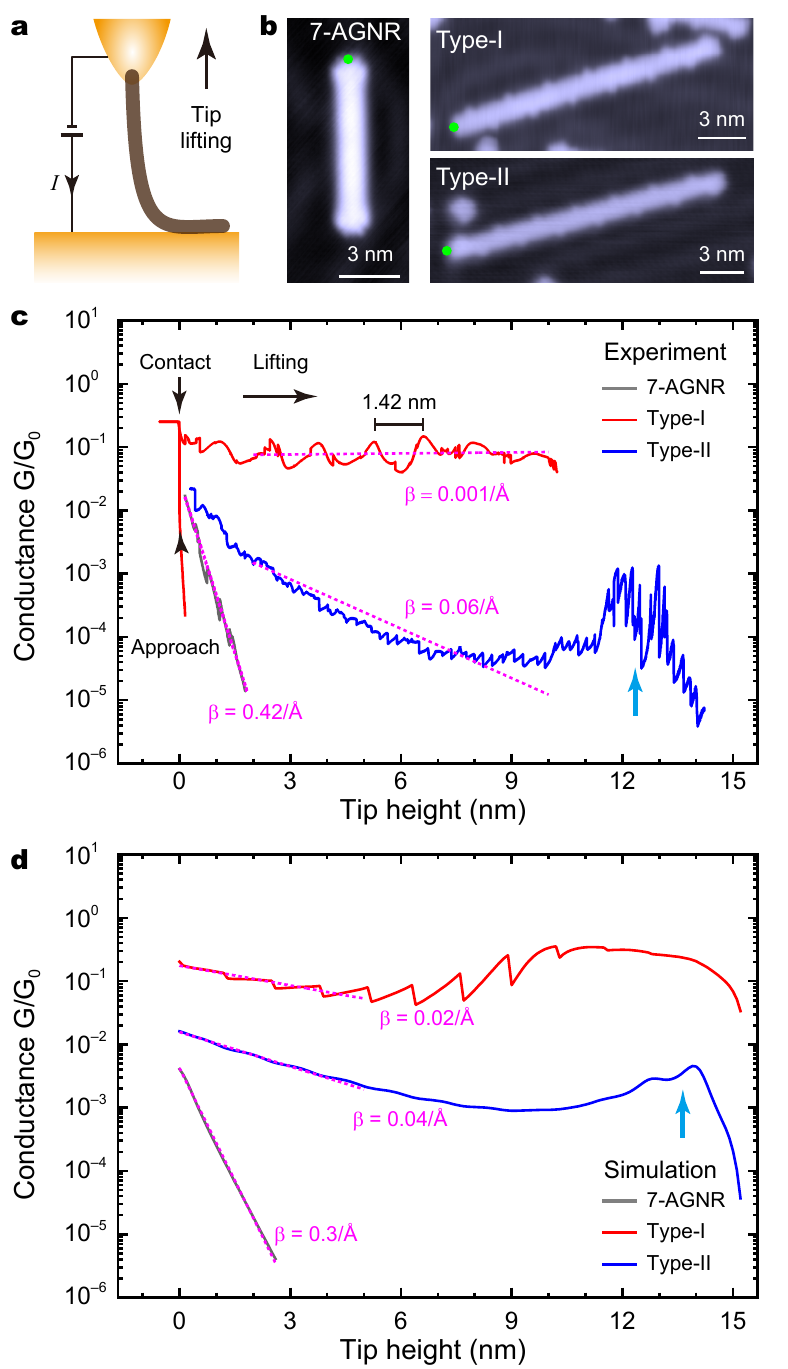}
\caption{\label{fig2}\textbf{Conductance measurements of suspended GNRs.} \textbf{a}, Schematic of the measurement setup during the lifting process. \textbf{b}, STM images of the three ribbons probed in \textbf{c}. STM image conditions: ($V = 50$ mV, $I = 10$ pA). \textbf{c}, Conductance measurements as a function of tip--sample separation for a 7-AGNR (grey curve) and two types of 7-AGNR-\textit{S}(1,3)s (red and blue curves). \textbf{d}, Characteristic conductance simulations for transport through lifted 7-AGNR and 7-AGNR-\textit{S}(1,3). Two different models for the valence band alignment with respect to the bias window captures qualitatively the observed difference between Type-I and II.}
\end{figure}

We now turn to the charge transport properties of both GNRs. The experimental procedure is sketched in Figure \ref{fig2}a. First, the STM tip is positioned above a terminus of a GNR with a low bias of $+50$ mV (unless noted otherwise), then the tip is approached until a contact with the GNR -- identified by a plateau in the conductance trace -- is established, and eventually the tip is lifted vertically from the surface with the GNR attached to the tip apex with the current traversing the junction recorded simultaneously. 

Typical conductance measurements for three OSS-fabricated GNRs, one 7-AGNR and two 7-AGNR-\textit{S}(1,3)s (STM images in Figure \ref{fig2}b), are shown in Figure \ref{fig2}c. For the 7-AGNR, the retraction of the tip reveals  an exponential conductance decrease (gray curve) with a decay constant of $\beta = 0.33$ {\AA$^{-1}$}, consistent with previous reports \cite{koch2012voltage}. 
Such a high $\beta$ value leads to a conductance lower than $10^{-5}$ G$_0$ for the ribbon lifted by only 2 nm, limiting the potential applications of 7-AGNRs in future electronic devices.
In contrast, for the 7-AGNR-\textit{S}(1,3)s we observed two distinct types of conductance behaviors as the tip is lifted, denoted Type-I (red) and Type-II (blue) in the following. Both types show remarkably high conductance at long lifting distances in comparison to that of 7-AGNRs.

The Type-I ribbon reported in Figure \ref{fig2}c (red curve) shows a near length-independent oscillatory conductance around a high value ($\sim 0.1$ G$_0$) with a periodicity of $\sim 1.42$ nm and an overall decay constant $\beta$ as low as $\sim 0.001$ {\AA$^{-1}$}. 
Some scattering around these values for other measured ribbons is reported in Supplementary
Information (SI) Section 1. 
The conductance oscillation has an average periodicity very close ($\sim 12$\% larger) to the distance of $1.3$ nm separating successive units of the 7-AGNR-\textit{S}(1,3) (Figure 1), suggesting that these conductance fluctuations reflect the detachment of successive units from the substrate surface \cite{Reecht2015}. 

The Type-II ribbon reported in Figure \ref{fig2}c (blue curve) follows a different behavior. 
The conductance decays smoothly over 6 to 7 nm before a plateau is reached at $G\sim 10^{-4}$ G$_0$, leading to a conductance a thousand times smaller than the similar Type-I ribbon for a lifted distance of 9 nm.
Surprisingly, the conductance of Type-II then \emph{increases} again upon further lifting up to $z >9$ nm (which would correspond to $\beta<0$). 
Eventually, the conductance decreases again for distances close to the overall length of the ribbon $z \sim 13$ nm.
The conductance traces also reveal tiny but abrupt jerks with a periodicity of $\sim 0.28$ nm (detailed in SI Section 1), which matches well with the lattice constant of the Au(111) surface.
A detailed analysis of STM images of several 7-AGNR-\textit{S}(1,3) of Type-I and Type-II does not reveal an obvious difference in their structures or their adsorption configurations on the substrate, hinting towards an effect resulting from different connections between the STM-tip and the GNR-terminus.

Transport simulations based on an effective TB model can capture the qualitative features of the experiments as shown in Figure \ref{fig2}d.
The 7-AGNR shows a characteristic decay of the conductance with suspended ribbon length due to the exponential decay of tunneling electrons in the energy range within the band gap.
The 7-AGNR-\textit{S}(1,3) on the other hand has a much smaller band gap which results in band states available much closer to the Fermi level $\varepsilon_F$ of the electrodes, and therefore to a higher overall conductance.
In fact, ballistic transport is obtained whenever a band state falls within the bias window.
In line with the general trend that GNRs are slightly $p$-doped on Au(111) \cite{Merino2017Width}, which is supported by our DFT calculations (SI Section 2), the VB states are the ones aligned closest to $\varepsilon_F$ and thus the ones involved in the electron transport process. 
Our model further enables us to reproduce the behaviour of both Type-I and Type-II ribbons by considering two different models for the molecular level alignment relative to $\varepsilon_F$ as elaborated below.

\begin{figure*}
\includegraphics[width=1\linewidth]{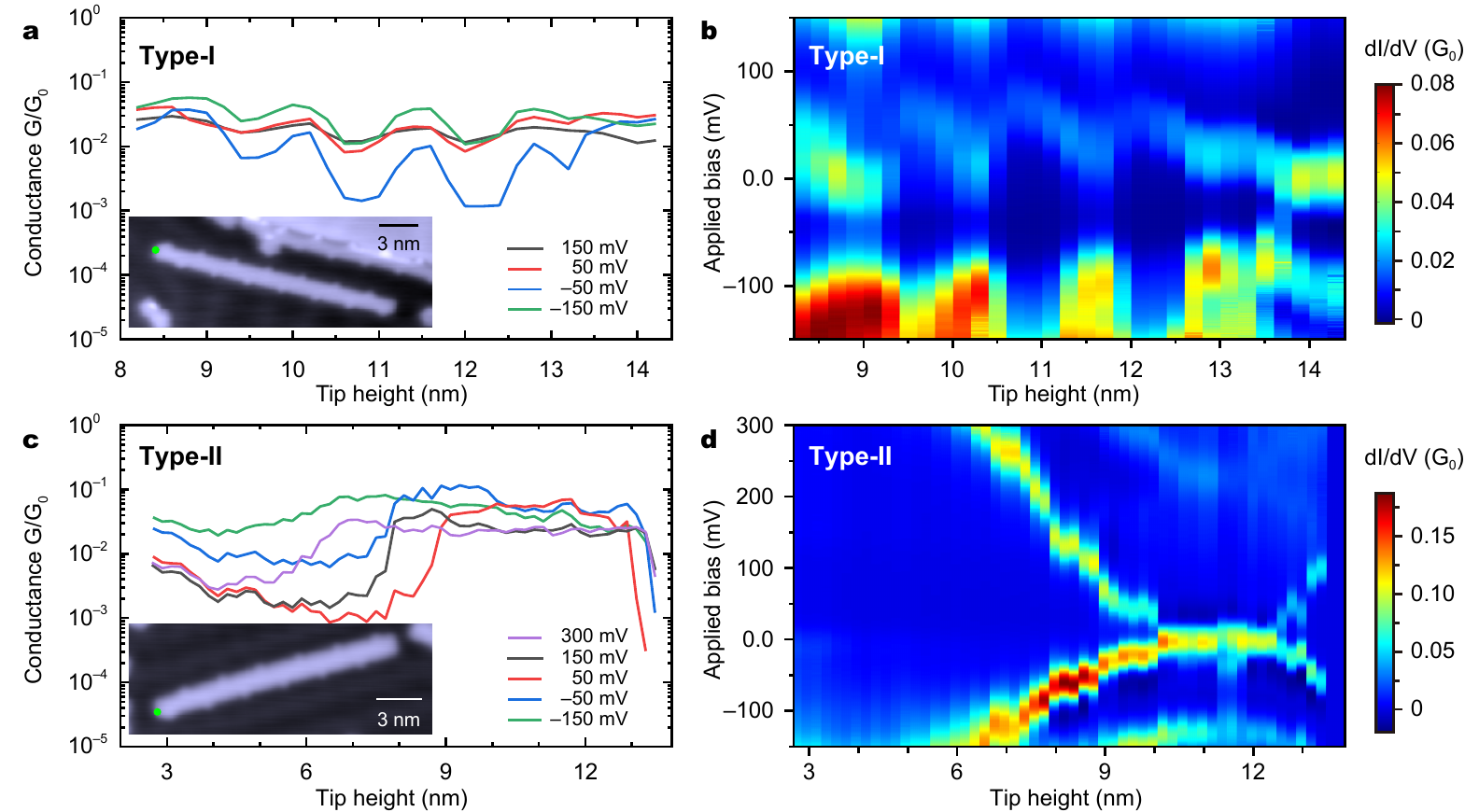}
\caption{\label{fig3}\textbf{Electronic transport mapping of two types of 7-AGNR-\textit{S}(1,3)}.
\textbf{a}, Bias-dependent conductance traces as a function of tip-sample separation for a Type-I ribbon reconstructed from the $I(V)$ spectra acquired at different tip height with a step separation of 0.2 nm. \textbf{b}, $dI/dV$ evolution acquired simultaneously to $I(V)$  in \textbf{a}. \textbf{c}-\textbf{d}, Corresponding measurements for a Type-II ribbon. Insets of \textbf{a} and \textbf{c} show the STM images ($V = 50$ mV, $I = 10$ pA) of the probed ribbons.
}
\end{figure*}

 In order to understand the physics behind these different transport characteristics, both conductance ($G=I/V$) and differential conductance ($dI/dV$) spectra were acquired during a step-by-step lifting procedure with 0.2 nm step size.
The conductance measurement of the Type-I ribbon (inset of Figure \ref{fig3}a) was performed with the tip retracting from 8.3 to 14.5 nm. Reconstructed from $I-V$ spectra acquired at discrete tip heights, the junction conductance at different voltages is represented in Figure \ref{fig3}a. Independently of the applied voltage, the junction conductance remains essentially constant around 0.02 G$_0$, with the exception of a characteristic oscillation with a periodicity of $\sim$ 1.4 nm already reported in Figure \ref{fig1}c.
The $dI/dV$ spectra acquired during the same lifting procedure (Figure \ref{fig3}b) reveal clear fingerprints of a transport resonance around the Fermi level and two others at $V\sim - 0.10$ V and $V \sim +0.15$ V. Whereas these three resonances show nearly constant intensities with tip retraction, their positions fluctuate with a periodicity that matches the 1.4 nm oscillations in the conductance trace (Figure \ref{fig3}a). A simple analysis of these spectra assigns the nearly length-independent conductance of Type-I ribbons to the presence of at least one of these resonances within the bias window, whereas the conductance oscillations seem due to shifts of the voltage resonances that appear with the periodicity of the edge modifications in the GNRs.

In Figure \ref{fig3}c and \ref{fig3}d we summarize the similar investigations with a Type-II ribbon. In agreement with the data presented in \Figref{fig2}c, the junction conductance is rather different from that of Type-I ribbons. Figure \ref{fig3}c reveals that, for all applied voltages, the junction conductance slowly decreases for short lifting distances before a sudden conductance increase is observed at lifting distances that depend on the applied voltage, \textit{i.e.,} the tip height onset occurs later at lower voltages. 
For short tip-sample separation ($z<5$ nm), $dI/dV$ spectra recorded simultaneously do not reveal any spectral features in the probed voltage window ($-0.15$ V $<V< +0.3$ V), in agreement with the exponentially decaying conductance observed at the early stage of the retraction for Type-II.    
When the ribbon is lifted further ($z>5$ nm), two resonances emerge (at $V= -0.1$ and 0.25 V for $z= 6.7$ nm). 
The voltage difference between these two peaks shrinks as the ribbon is further lifted, and eventually vanishes for $z\sim 10.5$ nm. For $z> 12$ nm the features split again, leaving the impression of two nearly parabolic dispersions. 
Beside these two spectral contributions, two other $dI/dV$ resonances briefly enter the bias windows (for 9 nm $<z<13$ nm) and follow a similar parabolic dispersion. 
From these data we conclude that the conductance decrease observed at low $z$-values corresponds to out-of-resonance transport conditions, whereas the conductance increase at $z \approx$ 6 to 9 nm, and the following conductance plateau, mark the onset of resonant transport through band states that have shifted into the bias window.

\begin{figure*}
\includegraphics[width=1\linewidth]{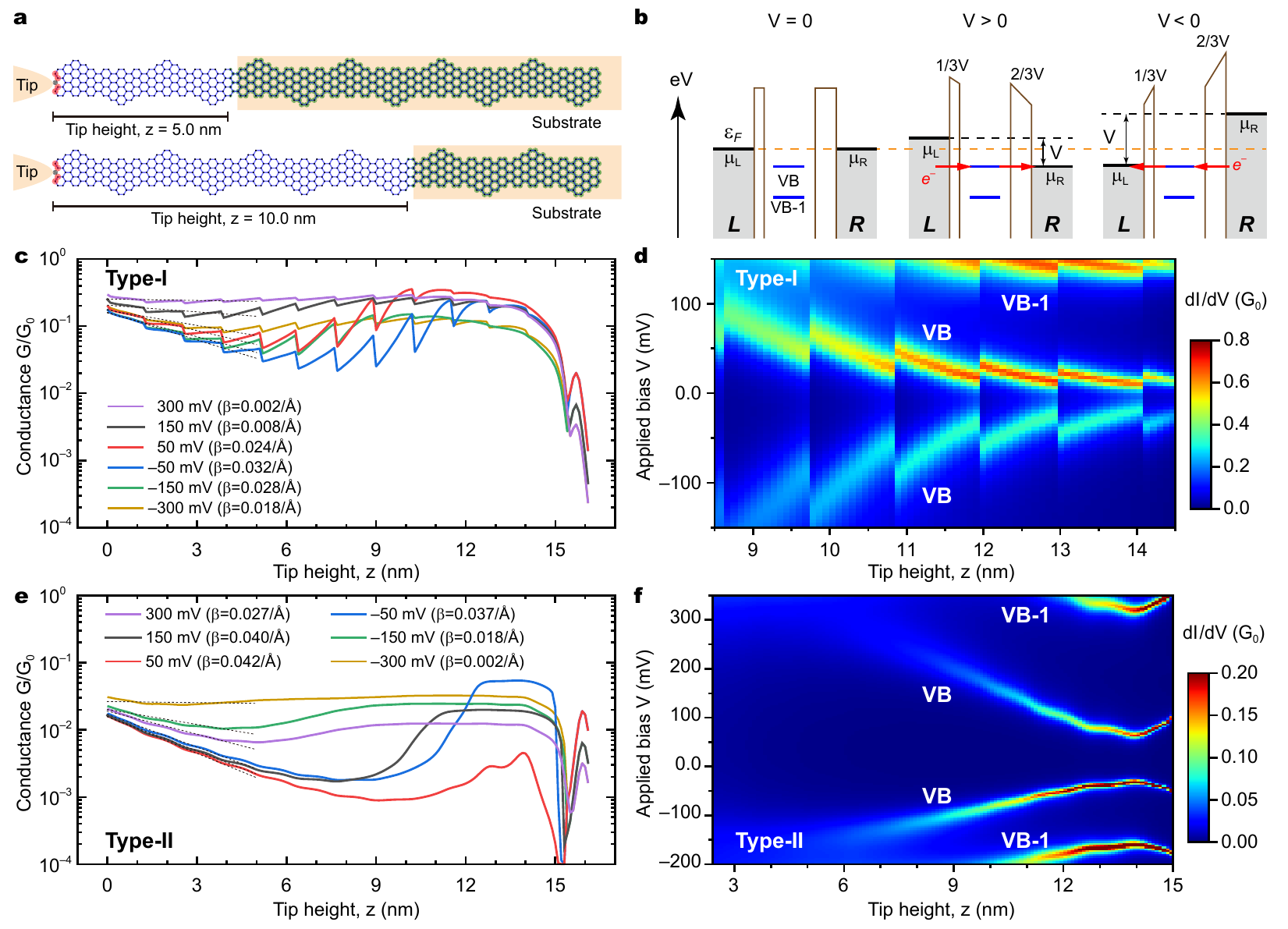}
\caption{\label{fig4}\textbf{Tight-binding model for transport through suspended GNRs}.
\textbf{a}, Top view of a 7-AGNR-\textit{S}(1,3) composed of $n=12$ precursor units corresponding to two different stages during the liftoff (suspended distances of $z=5.0$, $10.0$ nm, respectively). 
The vertices (black dots) indicate the carbon $2p_z$ orbitals in the hexagonal lattice. The apex atom of the STM tip is considered to form a chemical bond with the middle atom (gray disk) on the left zigzag edge.
The orbital sites indicated by red (green) circles indicate the relative strength of the tunneling rates to the tip (substrate) electrode.
\textbf{b}, Sketch of the energy-level alignments of the VB and VB-1 states in the junction,
relative to the Fermi level $\varepsilon_F$ (orange dashed line) under equilibrium conditions
and the chemical potentials $\mu_L$ and $\mu_R$ of the electrodes at different applied voltages:
$V = 0$ (left panel), $V > 0$ (middle panel) and $V < 0$ (right panel),
where $eV=\mu_L-\mu_R$.
\textbf{c}-\textbf{d}, Simulated conductance traces and $dI/dV$ for Type-I GNRs.
\textbf{e}-\textbf{f}, Simulated conductance traces and $dI/dV$ for Type-II GNRs.
}
\end{figure*}

The experimental trends reported in Figure \ref{fig3} are well captured by our model simulations.
Figure \ref{fig4}a illustrates the transport setup for two different tip heights. The left side of the ribbon is chemically bonded to the tip electrode at the center carbon atom (gray site) at the terminal zigzag edge and electrons are considered to tunnel (with rate $\Gamma_\mathrm{tip}$) between the tip electrode and the nearby $p_z$ orbitals (red sites) that form part of the conjugated structure.
Similarly, at the right side of the ribbon electrons are considered to tunnel (with rate $\Gamma_\mathrm{sub}$) between the substrate electrode and the $p_z$ orbitals directly above it (green sites).
In other words, the tip height is modeled by varying the number of sites in contact with the substrate, with direct consequences for the electron transport between the electrodes.

Besides the tunnel couplings $\Gamma_\mathrm{tip}$ and $\Gamma_\mathrm{sub}$, two more parameters are included in our simulation: the charge state of the GNR and the ratio of the voltage drops at the two electrode interfaces (tip-GNR and GNR-Au(111)).
First, as sketched in Figure \ref{fig4}b, in equilibrium the VB and VB-1 states are aligned just below the Fermi level $\varepsilon_F$, consistent with the trend that GNRs on Au(111) typically result in $p$-type doping \cite{Merino2017Width}. Such level alignment is also corroborated by our DFT calculations reported in the SI Section 2.
Secondly, when a voltage is applied, our model assumes an asymmetric potential profile with voltage drops of 1/3 and 2/3 over the two interfaces, respectively.
A voltage drop over both interfaces implies that the VB resonances are seen in $dI/dV$ for both bias polarities \cite{Nazin2005Tunneling}.

In essence, length-independent transport is therefore observed whenever a VB state is accessible within the energy window given by the applied bias voltage (Figure \ref{fig4}b).
Conversely, an exponentially decaying conductance trace is observed while the GNR states fall outside the bias window. As explained in SI Section 3, our interpretation of the difference between the experimentally observed Type-I and Type-II ribbons relies on the following physical mechanisms:
(i) Type-I ribbons are considered to be strongly bound to the tip electrode such that most of the applied voltage drop occurs over the GNR-substrate interface (i.e., $L$ is tip and $R$ is substrate in Figure \ref{fig4}b). Further, the precise $p$-doping level of the GNR is considered to be correlated with the detachment of each precursor unit. A possible origin of this may be related to subtle modifications of the GNR/substrate interface during the detachment, which in turn leads to conductance oscillations with a periodicity of $\sim 1.4$ nm.
(ii) Type-II ribbons are considered to be weaker bound to the tip electrode such that most of the voltage drop occurs over the tip-GNR interface (\textit{i.e.}, $L$ is substrate and $R$ is tip in Figure \ref{fig4}b).
The donation of charge is not considered to be significantly correlated with the liftoff process for Type-II, hence the absence of the conductance oscillations with a periodicity of $\sim 1.4$ nm as observed for Type-I.
These considerations can qualitatively explain the experimental data as explained in the following.

Figure \ref{fig4}c and \ref{fig4}d report the simulated conductance $G=I/V$ and the differential conductance $dI/dV$ at different voltages for Type-I ribbons.
Overall, the traces are characterized by length-independent transport from start to the end.
At positive polarity the conductance is somewhat larger and the plateau more flat than at negative polarity. As in the experiment (Figure \ref{fig3}a), the oscillations are most pronounced at small voltages.
The features observed in $dI/dV$ are fingerprints of the VB and VB-1 states that appear at both polarities (Figure \ref{fig4}b) due to voltage drops over both interfaces.
The spacing is closer at the positive polarity because most of the voltage drop is assumed to occur over the GNR-substrate interface.

Figure \ref{fig4}e and \ref{fig4}f for Type-II ribbons, on the other hand, exhibit the initial phase of the liftoff process that is characterized by an exponential decrease of conductance due to the VB states aligned outside the transport bias window.
However, as the GNR is gradually removed from the substrate, the molecular resonances narrow in energy (reduced substrate hybridization) and, consistent with our assumption of a constant $p$-doping level, must therefore drift up in energy.
This eventually brings the VB and VB-1 states inside the bias window and the conductance trace \emph{increases} to reach a high-conductance plateau similar to Type-I.
Again, the features observed in $dI/dV$ are fingerprints of the VB and VB-1 resonances that appear at both polarities due to a voltage drop over both interfaces.
But contrary to Type-I, the spacing is now closer at the negative polarity because most of the voltage drop is assumed to occur over the tip-GNR interface.

In conclusion, we performed a systematic study of the transport properties of the edge-modified 7-AGNR-\textit{S}(1,3) in a LT-STM liftoff procedure. These ribbons host electronic low-energy topological bands supported by zigzag-edge decorations. These topological bands are responsible for length-independent and efficient charge transport in the suspended GNR geometry. The transport behavior is highly sensitive to the coupling between the ribbon and the electrodes.
Overall, the critical factor for high-conductance ballistic transport is related to the presence of delocalized valence band states within the bias window. Future studies may explore further the possible relationship between the topological nature of the band and the extraordinary transport characteristics of the 7-AGNR-\textit{S}(1,3). Our approach opens a path to the realization of tunable molecular components based on the rational design of graphene nanoribbons edges.

\textbf{Acknowledgements}
This project has received funding from the European Research Council (ERC) under the European Union's Horizon 2020 research and innovation program (grant agreement No 771850). The International Center for Frontier Research in Chemistry (FRC) is also acknowledged for financial support. 
This research was supported by the Swiss National Science Foundation under Grant No. 200020{\_}182015, the Werner Siemens Foundation (CarboQuant).
TF acknowledges support by the Spanish MCIN/AEI/
10.13039/501100011033 (PID2020-115406GB-I00), 
the Basque Department of Education (PIBA-2020-1-0014), and the European Union through Horizon 2020 (FET-Open project ``{SPRING}'' Grant no.~863098).K.M. acknowledges a fellowship from Gutenberg Research College Johannes Gutenberg University Mainz. R.F. thanks Oliver Gröning for stimulating discussions.

\section{Methods}
The experiments were performed with a low-temperature ultrahigh-vacuum STM (Omicron) at a base pressure of $\sim 5 \times 10^{-11}$ torr at 4.6 K. 
Atomically flat Au(111) substrates were prepared via cycles of argon ion sputtering and annealing under ultrahigh vacuum conditions.
Electrochemically etched W tips were covered with gold by repeated indentation into the substrate. 
On-surface synthesis of these GNRs was performed using 10,10'-dibromo-9,9'-bianthryl (DBBA, Figure \ref{fig1}a) \cite{Cai2010} and  6,11-bis(10-bromoanthracen-9-yl)-1,4-dimethyltetracene (BADMT, Figure \ref{fig1}b) \cite{groning2018engineering} as precursors for 7-AGNRs and  7-AGNR-\textit{S}(1,3)s, respectively. The precursors were sublimed onto freshly cleaned Au(111) substrates held at room temperature. The samples were then gradually annealed to 400 {\textcelsius}, leading to the formation of the GNRs as detailed in previous studies \cite{Cai2010, groning2018engineering}. Differential conductance d$I$/d$V$ spectra were recorded in constant-height mode (open feedback loop) with a voltage modulation of 20 mV and a modulation frequency of 901 Hz.

To simulate the electronic properties of suspended GNRs we adopted a transparent and simple description based on a nearest-neighbor TB model of the carbon $p_z$ orbitals with hopping matrix elements set to $|t|=3$ eV \cite{zerothi_sisl}.
The low-bias differential conductance $dI/dV$ as a function of the liftoff distance was simulated using Green's function theory in the wide-band approximation for the tip and substrate electrodes as described in the SI. We ignore effects of curvature by considering a model as illustrated in Figure \ref{fig4}a. The tip apex atom is considered to form a strong chemical bond to the central carbon atom at the terminal zigzag edge, resulting in $sp^3$ hybridization and thus in removal of the corresponding $p_z$ orbital from the conjugated network. The tip electrode is consequently considered to inject/extract electrons to/from the closest $sp^2$ carbon sites (red disks) characterized by the tunneling rate $\Gamma_\mathrm{tip}$. On the other hand, the number of atomic sites with coupling to the substrate electrode $\Gamma_\mathrm{sub}$ (green disks) is considered to be gradually reduced as a function of tip retraction from the surface.
This numerically efficient description allowed us to explore a wide range of transport scenarios and possible effects, including different models for the potential drop across the GNR, deviations from charge neutrality, disorder, etc. Based on this exploration we found that the most suitable description for our experimental observations corresponds to a model in which the Fermi level $\varepsilon_F$ is aligned close to the VB edge and where the current---at both bias polarities---involves propagation through a few VB states that fall within the transport energy window. Another key ingredient is the precise molecular level alignment relative to $\varepsilon_F$, which our modeling suggests to undergo small shifts as a function of the tip height due to changes in the electrode couplings. This enables a switching of the VB states in or out of the transport window.
The key parameters and assumptions in our transport model are corroborated by DFT calculations for the electronic structure of a partly lifted 7-AGNR-\textit{S}(1,3) on Au(111) as explained in the SI Section 2.

\clearpage
\bibliographystyle{apsrev-title}
\bibliography{Reference.bib}
\end{document}